\documentclass[aps,pra,twocolumn,floatfix,showpacs]{revtex4}
\usepackage{graphicx}

\begin{document}

\title{Strong spin-oscillation of small spin-1 condensates caused by an inclined
weak magnetic field}

\author{Y. Z. He}
\author{Z. F. Chen}
\author{Z. B. Li}
\author{C. G. Bao}
\thanks{The corresponding author}
\affiliation{State Key Laboratory of Optoelectronic Materials and
Technologies, School of Physics and Engineering, Sun Yat-Sen
University, Guangzhou, 510275, P.R. China}

\begin{abstract}
When a magnetic field is applied along a direction deviated from the
quantization $Z$-axis, the conservation of total magnetization holds
no more. In this case the inclined field can cause a strong
spin-evolution via the linear Zeeman term even the field is as weak
as a percentage of $mG$. An approach beyond the mean field theory is
proposed to study the evolution of small $^{87}$Rb condensates under
the weak inclined fields. The time-dependent populations of
spin-components are given in analytical forms. The evolution is
found to be highly sensitive to the magnitude and direction of the
field.
\end{abstract}

\pacs{
      03.75.Mn, % Multicomponent condensates; spinor condensates
      03.75.Kk  % Dynamic properties of condensates; collective and hydrodynamic excitations, superfluid flow
}

\maketitle

\section{Introduction}

Since the experimental realization of spinor Bose-Einstein
condensates\cite{ho98,ohmi98,stam98,sten98,goel03,grie05}, the
spin-evolution of the condensates has become a hot topic due to its
academic interest and potential application. The manipulation of the
evolution is a central
problem\cite{sore01,chang2004,youli2005,law98,pu99,chan07,
uchi08,cui2008,jing08,mark08}. Usually, one applies an external
magnetic field lying along the direction of the axis of quantization
($Z$-axis)\cite{chang2004,kuwa04,mur06,kron06}. In this way the
total magnetization of the condensate is conserved, hence the linear
Zeeman term of the field plays no role. However, if the direction of
the field deviates from the $Z$-axis, the conservation of the total
magnetization will not exist. Accordingly, linear Zeeman term will
affect the evolution. Since the linear term is much stronger than
the quadratic term, a very weak inclined magnetic field might cause
remarkable effect. The dynamic response of the condensate to a
transversal magnetic field has been studied by Yi and
Pu\cite{pu00,yi06}. Strong oscillation has been found in the
evolution of populations of spin-components. Spin squeezing and
macroscopic entanglement have been found in the studies of the
ground state structures. Since the direction of the field might
affect the dynamic phenomena of condensates sensitively, this topic
deserves to be further studied.

The aim of this paper is to study the effect of an inclined magnetic
field (lying along an arbitrary $Z^{\prime}$-axis) on the
spin-evolution of the condensates of $^{87}$Rb atoms. Recently, more
attention was paid to the study of small condensates with lower
densities because the dissipative process can be reduced, and
therefore richer phenomena might be observed\cite{zhan05}. For these
small systems the validity of the mean field theory might be
limited. Therefore, in this paper, an approach beyond the mean field
theory is proposed. It turns out that analytical solutions can be
obtained as follows.

\section{One-body system}

In order to understand better the effect of the inclined $B$ on
many-body systems, we study firstly a simple example, namely, the
evolution of a single spin-1 $^{87}$Rb atom under the field. The
initial spin-state of the atom $\chi_{\mu}$ is defined in a
$Z$-frame, while $B$ is lying along another axis $Z^{\prime}$. The
angle between $Z$ and $Z^{\prime}$ is $\theta$. The Hamiltonian
$H=-p\hat{S}_{Z^{\prime}}$, where $p=\gamma B$, and $\gamma
=g_{F}\mu_{B}$ being the gyromagnetic ratio. The quadratic Zeeman
term is much weaker than the linear term (e.g., if $B=1\ mG$, the
former is $10^{-7}$ times weaker). Therefore, when $Z^{\prime}$ and
$Z$ do not overlap, the former can be neglected.

Due to the inclined magnetic field the spin-state at time $t$
becomes
\begin{equation}
 \xi _{\mu }(t)
 =e^{-iHt/\hbar }\chi _{\mu }
 =\sum_{\nu} d_{\nu \mu}^{1}(-\theta ) e^{i\nu \tau}
  \chi _{\nu}^{\prime}
 \label{e1}
\end{equation}
where $\tau=pt/\hbar$, $\chi_{\nu}^{\prime}$ is defined in the
$Z^{\prime}$-frame, and $d_{\nu \mu}^{1}(-\theta)$ is an element of
the well-known rotation matrix. Since the observation is made in the
$Z$-frame, Eq.~(\ref{e1}) is rewritten as
\begin{equation}
 \xi _{\mu }(t)
 =\sum_{\lambda } M_{\mu \lambda}(t) \chi_{\lambda}
 \label{e2}
\end{equation}
where
\begin{equation}
 M_{\mu \lambda}(t)
 =\sum_{\nu} d_{\nu \mu }^{1}(-\theta) e^{i\nu\tau}
  d_{\lambda \nu}^{1}(\theta)
\end{equation}

It is obvious that $|M_{\mu \lambda}(t)|^{2} \equiv
P_{\lambda}^{\mu}(t)$ is the probability that an atom in $\mu$
initially would be in $\lambda$ at $t$ due to the inclined $B$. The
evolution appearing as a variation of $P_{\lambda}^{\mu}(t)$ is
strictly periodic with the period $\tau = 2\pi$ or $t=h/p \equiv
t_{p}$. For an example, when $B=1\ mG$, $t_{p}=1.41\ ms$.

It is obvious that $\{M_{\mu \lambda}\}$ is a unitary matrix and is
symmetric. Due to the symmetry of $d_{\nu \mu}^{1}$, we have
\begin{equation}
 M_{\mu \lambda}
 =M_{\lambda \mu}
 =(-1)^{\mu+\lambda } M_{-\mu,-\lambda }^{\ast}
 \label{e3}
\end{equation}

In particular,
\begin{equation}
 M_{1,1}
 =\frac{1+\cos^{2}\theta}{2} \cos(\tau )
  +\frac{\sin^{2}\theta }{2}
  +i\cos\theta\sin (\tau)
 \label{e4_M11}
\end{equation}
\begin{equation}
 M_{1,0}
 =\frac{\cos\theta \sin\theta }{\sqrt{2}}(\cos(\tau)-1)
  +i\frac{\sin\theta}{\sqrt{2}}\sin(\tau)
\end{equation}
\begin{equation}
 M_{1,-1}
 =\frac{\sin^{2}\theta}{2}(\cos(\tau )-1)
\end{equation}
\begin{equation}
 M_{0,0}
 =\sin^{2}\theta \cos(\tau )
 +\cos^{2}\theta
\end{equation}

The other elements of $\{M_{\mu \lambda}\}$ can be obtained via
Eq.~(\ref{e3}). Thus the evolution is completely clear.

Incidentally, based on the mean field theory, the effect of an
inclined magnetic field has been studied by Pu, et al\cite{pu00}.
When the atom-atom interaction and the quadratic Zeeman term have
been neglected, they have derived a set of dynamic equations for the
time evolution of the field amplitudes $a_{\lambda}(t)$.  When the
condensate is initially fully polarized ($a_{1}(0)=1$ and
$a_{0}(0)=a_{-1}(0)=0$), the set $a_{\lambda}(t)$ has an analytical
solution as shown by Eq.~(\ref{e4_M11}) of their paper. It turns out
$a_{\lambda}(t)=M_{1,\lambda}$. Thus, for this case, the mean field
theory for many-body systems and the above simple consideration for
a single-body system lead to the same result.

\section{Many-body systems without atom-atom interaction}

The initial state is assumed to be a Fock-state defined in the
Z-frame as $|I\rangle \equiv
|N_{1}^{I},N_{0}^{I},N_{-1}^{I}\rangle$, where $N_{\mu}^{I}$ is the
number of atoms in $\chi_{\mu}$ initially. When the atom-atom
interaction is neglected, it is straight forward to obtain the
probability of an atom in $\lambda$ as
\begin{equation}
 P_{\lambda}^{I,\theta}(t)
 =\sum_{\nu} |M_{\nu \lambda}|^{2} N^{I}_{\nu}/N
 \label{e4}
\end{equation}

As before, the observation is made in the $Z$-frame. A notable point
is that $P_{\lambda}^{I,\theta}$ does not depend on $N$ but the
ratio $N_{\nu}/N$, and it has the same period $h/p$ as the single
atom has. Furthermore, it is invariant under a reflection against
the $X$-$Y$ plane, $P_{\lambda}^{I,\theta} =
P_{\lambda}^{I,\pi-\theta}$. In particular, for $\lambda=0$
\begin{equation}
 P_{0}^{I,\theta }
 =X(1-3\frac{N_{0}^{I}}{N})
  +\frac{N_{0}^{I}}{N}
 \label{e5}
\end{equation}
where
\begin{equation}
 X
 =\frac{\sin^{2}\theta}{2}
  \{1-\sin^{2}\theta \cos^{2}(\tau)
   +\cos^{2}\theta [1-2\cos(\tau)] \}
\end{equation}

Eq.~(\ref{e5}) implies that the probability is not at all affected
by the initial magnetization $N_{1}^{I}-N_{-1}^{I} \equiv M^{I}$,
but is seriously affected by the number of atoms initially in
$\mu=0$. From Eq.~(\ref{e5}), we know that whether $N_{0}^{I}/N$ is
larger or smaller than $1/3$ is crucial to the evolution. If
$N_{0}^{I}=N/3$, $P_{0}^{I,\theta }$ would remain constant (without
evolution). Otherwise, $P_{0}^{I,\theta }$ will oscillate around a
background, and the amplitude would become the largest if
$N_{0}^{I}=N$. On the other hand, if $Z^{\prime}$ and $Z$ overlaps,
we have $X=0$ and $ P_{0}^{I,\theta }$ remains also constant as
expected.

The time-dependent magnetization $N(P_{1}^{I,\theta} -
P_{-1}^{I,\theta}) \equiv N P_{mag}^{I,\theta}$ can be obtained from
Eq.~(\ref{e4}), we have
\begin{equation}
 P_{mag}^{I,\theta }(t)
 =M^{I} [\cos^{2}\theta + \sin^{2}\theta \cos(\tau)]/N
 \label{e6}
\end{equation}

It implies that the time-dependent magnetization depends on the
initial magnetization $M^{I}$ but is not affected by $N_{0}^{I}/N$.
In particular, if the system is zero-polarized initially,
$P_{mag}^{I,\theta}(t)$ remains zero (without evolution). On the
other hand, when $Z^{\prime}$ and $Z$ overlaps, $P_{mag}^{I,0}(t)$
remains to be a constant $M^{I}/N$ as expected.

Incidentally, when $\theta = \pi /4$, the evolution of
$P_{\lambda}^{I,\theta}(t)$ has been calculated numerically in
\cite{pu00} and plotted in Fig.3 of their paper. We found that the
difference between their numerical results and those from
Eq.(\ref{e4}) is very small. The small difference implies that, in
the early stage of evolution, the strong oscillation shown in their
figure is essentially caused by the inclined field and is less
affected by the interaction.

\section{Many-body systems with atom-atom interaction}

For realistic condensates of $^{87}$Rb atoms as an example, the
interaction
\begin{equation}
 v_{ij}
 =\delta(\mathbf{r}_{i}-\mathbf{r}_{j})
  \sum_{s} g_{s} \mathfrak{P}_{ij}^{s}
 \label{e7}
\end{equation}
where the strength $g_{s}=4\pi \hbar^{2}a_{s}/M$, $M$ is the mass of
atom. $\mathfrak{P}_{ij}^{s}$ is the projection operator of the
$s$-spin-channel ($s=0$ or $2$, which is the total spin of the two
atoms $i$ and $j$). $a_{0}=101.8a_{B}$, $a_{2}=100.4a_{B}$ from
\cite{kemp99}. It is assumed that the number density of the
condensate and the temperature are sufficiently low so that the
single-spatial-mode approximation can be adopted\cite{zhan05}. Under
this approximation, when an irrelevant constant and the quadratic
Zeeman term have been dropped, the Hamiltonian reads \cite{law98}
\begin{equation}
 H
 =G \hat{S}^{2}
  -p\sum_{i} \hat{S}_{Z^{\prime}i}
 \label{e8}
\end{equation}
where $\hat{S}$ is the operator of the total spin of the many-body
system, $G=\frac{1}{6} (g_{2}-g_{0}) \int
d\mathbf{r}|\phi(\mathbf{r})|^{4}$, $\phi(\mathbf{r)}$ is the
spatial normalized wave function of an atom (all atoms are assumed
to condense into this state). Since the details of
$\phi(\mathbf{r)}$ affects only the strength $G$, it is not
essential to our qualitative results. Therefore, it is simply
evaluated via the Thomas-Fermi approximation.

Let us introduce the total spin-state $\vartheta_{S,M}$ of the whole
system with conserved total spin $S$ and its $Z$-component $M$. The
overlap of this state and the initial state $\langle \vartheta_{S,M}
| N_{1}^{I},N_{0}^{I},N_{-1}^{I} \rangle \equiv \delta_{M,M^{I}}
D_{M^{I},N_{0}^{I}}^{N,S}$ is essential to the following
calculation. These coefficients have been given explicitly by Wu
(Eqs.~(36) and (37) of \cite{wu96}). They could be derived also by a
set of recursion formulae\cite{luo08a}. On the other hand, we define
further the total spin states $\vartheta_{S,M^{\prime}}^{\prime}$
relative to the $Z^{\prime}$-frame,
$\vartheta_{S,M^{\prime}}^{\prime}$ is related to $\vartheta_{S,M}$
via a rotation. Thereby the initial state defined in the $Z$-frame
can be expanded by $\vartheta_{S,M^{\prime}}^{\prime}$ as
\begin{equation}
 |I\rangle
 =\sum_{S,M^{\prime}} C_{S,M^{\prime}}^{I,\theta}
  \vartheta_{S,M^{\prime}}^{\prime}
\end{equation}
where $C_{S,M^{\prime}}^{I,\theta}
=D_{M^{I},N_{0}^{I}}^{N,S}d_{M^{\prime},M^{I}}^{S}(-\theta)$

Since the set $\vartheta_{S,M^{\prime}}^{\prime}$ are the
eigen-states of the Hamiltonian with the eigen-energy
$GS(S+1)-pM^{^{\prime}}$, the time-dependent solution of the system
reads
\begin{equation}
 \Psi (t)
 =\sum_{S,M^{\prime}} C_{S,M^{\prime}}^{I,\theta}
  e^{-i(GS(S+1)/p-M^{\prime})\tau}
  \vartheta_{S,M^{\prime}}^{\prime}
 \label{e9}
\end{equation}

By using the fractional parentage coefficients given in
\cite{bao04,bao05}, we can extract the spin-state of a single
particle (say, particle 1) from $\vartheta _{S,M^{\prime}}^{\prime}$
as
\begin{eqnarray}
 \vartheta_{S,M^{\prime }}^{\prime }
 &=&\sum_{\mu} \chi_{\mu}^{\prime}(1)
  [\mathcal{A}(N,S,M^{\prime},\mu)
   \vartheta_{S+1,M^{\prime}-\mu}^{[N-1]\prime}  \nonumber \\
 & &+\mathcal{B}(N,S,M^{\prime},\mu)
  \vartheta_{S-1,M^{\prime}-\mu}^{[N-1]\prime}]
 \label{e10}
\end{eqnarray}
where
\begin{equation}
 \mathcal{A}(N,S,M^{\prime },\mu)
 =[\frac{(N-S)(S+1)}{N(2S+1)}]^{1/2}
  C_{S+1,M^{\prime}-\mu,1,\mu }^{S,M^{\prime}}
 \label{A1}
\end{equation}
\begin{equation}
 \mathcal{B}(N,S,M^{\prime },\mu)
 =[\frac{S(N+S+1)}{N(2S+1)}]^{1/2}
  C_{S-1,M^{\prime}-\mu,1,\mu}^{S,M^{\prime}}
 \label{B1}
\end{equation}
where the Clebsch-Gordan coefficients have been introduced. Note
that in Eqs.~(\ref{A1}) and (\ref{B1}) $N-S$ must be even, otherwise
the state $\vartheta_{S,M^{\prime}}^{\prime}$ does not exist.

Since the observation is made in the $Z$-frame, the single particle
state $\chi_{\mu}^{\prime}(1)$ in Eq.~(\ref{e10}) is further
rewritten as $\chi_{\mu}^{\prime}=\sum_{\lambda} d_{\lambda
,\mu}^{1}(\theta) \chi_{\lambda}$, where $\chi_{\lambda}$ is defined
in the $Z$-frame.

With these transformations, eventually the probability of a particle
in $\lambda$ can be extracted from $\Psi(t)$, and we have
\begin{eqnarray}
 &&P_{\lambda}^{I,\theta}(t)
  = \sum_{\mu^{\prime},\mu } d_{\lambda,\mu^{\prime}}^{1}(\theta)
  d_{\lambda,\mu}^{1}(\theta)
  \sum_{S,M^{^{\prime \prime}},M^{\prime}}
   \delta_{M^{\prime \prime}-\mu^{\prime},M^{\prime}-\mu}  \nonumber \\
 & &\ \ \ \{ C_{S,M^{\prime \prime}}^{I,\theta}
  C_{S,M^{\prime}}^{I,\theta}
  [ \mathcal{A}(NSM^{\prime \prime} \mu^{\prime})
   \mathcal{A}(NSM^{\prime}\mu)  \nonumber \\
 & &\ \ \ +\mathcal{B}(NSM^{\prime \prime} \mu^{\prime})
  \mathcal{B}(NSM^{\prime}\mu)]
  \cos((M^{\prime \prime}-M^{\prime})\tau)  \nonumber \\
 & &\ \ \ +2C_{S+2,M^{\prime \prime}}^{I,\theta}
  C_{S,M^{\prime}}^{I,\theta}
  \mathcal{B}(N,S+2,M^{\prime \prime} \mu^{\prime})
  \mathcal{A}(NSM^{\prime}\mu)  \nonumber \\
 & &\ \ \ \cos([4G(S+3/2)/p-(M^{\prime \prime}-M^{\prime })]\tau)\ \}
 \label{e11}
\end{eqnarray}

This is a generalized version of Eq.~(\ref{e4}) with the realistic
interaction taken into account. It is also a generalization of
Eq.~(3) of \cite{luo08b}, taking the effect of the inclined magnetic
field into account. When $\theta=0$, Eq.~(\ref{e11}) is identical to
Eq.~(3) of \cite{luo08b}.

Eq.~(\ref{e11}) provides an analytical description of the evolution
and all the coefficients involved have analytical forms. Based on
Eq.~(\ref{e11}), numerical results are shown by the following
figures as examples to demonstrate the feature of evolution. The
condensate of $^{87}$Rb atoms is assumed to be trapped by a harmonic
potential with $\omega =300\times 2\pi$. A very weak field with
$B=0.01\ mG$ is chosen. This is enough to show the effect of the
linear Zeeman term. Although Eq.~(\ref{e11}) holds for arbitrary
$N$, numerical results are limited by the ability of computer. As
the first example, $N=100$ is chosen.

\begin{figure}[tbp]
 \centering
 \resizebox{0.99\columnwidth}{!}{ \includegraphics{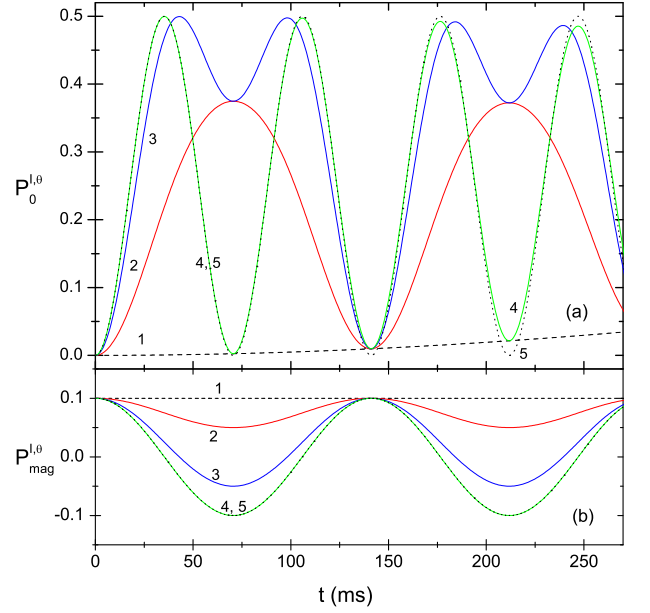} }
 \caption{(Color online) $P_{0}^{I,\theta}(t)$ (a) and $P_{mag}^{I,
\theta}(t)\equiv P_{1}^{I,\theta}-P_{-1}^{I,\theta}$ (b) against $t$
calculated from Eq.~(\protect\ref{e11}) with $\omega =300\times 2\pi
$, $N=100$ and $B=0.01mG$. The initial state has $M^{I}/N=0.1$ and
$N_{0}^{I}=0$. For the curves ``1" to ``5", $\theta =0$, $\pi /6$,
$\pi /3$, $\pi /2$, and $\pi /2$, respectively. $P_{\lambda}^{I,\pi
-\theta}(t)=P_{\lambda}^{I,\theta}(t)$ holds always. The curve ``1"
is identical to the one with $B=0$. ``5" is for the case with the
interaction ignored ($G=0$).}
 \label{F1}
\end{figure}

\begin{figure}[tbp]
 \centering
 \resizebox{0.99\columnwidth}{!}{ \includegraphics{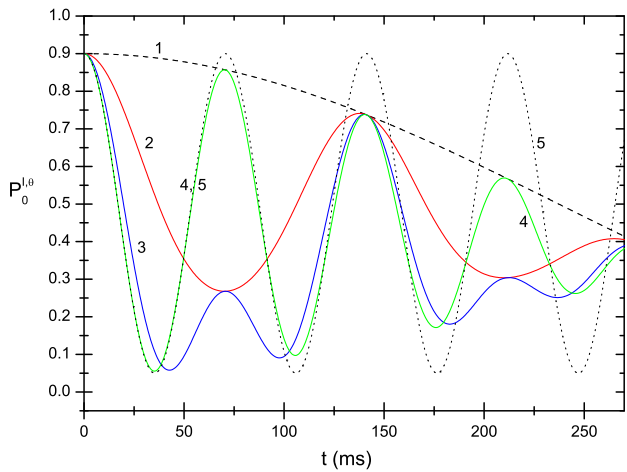} }
 \caption{(Color online) All the same as Fig.~\ref{F1}a but
with $N_{0}^{I}=0.9$.}
 \label{F2}
\end{figure}

In Figs.~\ref{F1} and \ref{F2}, $\theta $ is given at a set of
values. Related to the figures the following points are noted.

(i) Comparing curves ``2" to ``4" (in solid line) with ``1" (in dash
line), the strong and swift oscillation caused by the inclined weak
magnetic field is impressive. The evolution is highly sensitive to
$\theta$.

(ii) When the interaction is ignored, the evolution is described by
the curve ``5" (in dotted line) which arises purely from the
inclined field. Comparing ``4" and ``5" (both have $\theta =\pi
/2$), we know that the effect of interaction is weak in the early
stage because the two curves almost overlap. However, if $t$ is
larger, the influence of interaction would become more and more
explicit, and the deviation between ``4" and ``5" would be serious
(refer to Fig.~\ref{F2}).

(iii) The evolutions shown by the curves (except ``1") are nearly
periodic with the period $t_{p}=h/p=141\ ms$. If the interaction is
removed or if $G/p$ is an integer, it would be exactly periodic as
shown by Eq.~(\ref{e11}). However, since $t_{p}$ is reversely
proportional to $B$, the cycles of oscillation would become very
dense if $B$ is large. For instance, when $B$ $\geq 1\ mG$, the
curves of $P_{\lambda}^{I,\theta}(t)$ with $\theta \neq 0$ look like
a band. The width of the band is equal to a double of the amplitude
of oscillation, and therefore depends on $\theta$. However, if
$\theta$ is not small (say, $\theta \geq \pi /15$), the width would
be too broad and the band is difficult to be determined. Thus, when
a stronger field is used in experiments, only the cases with small
$\theta$ are meaningful.

(iv) When the field is not inclined, magnetization is conserved as
shown by ``1" of Fig.~\ref{F1}b. The inclination breaks the
conservation and causes oscillation as shown by ``2" to ``4" of
Fig.~\ref{F1}b. The oscillation would have the largest amplitude if
$\theta =\pi /2$. Meanwhile, $P_{mag}^{I,\pi /2}(t)$ oscillates
between $\pm M^{I}/N$. It implies that, when the initial
magnetization is larger, the amplitude is larger. The overlap of
``4" and ``5" in 1b implies that the interaction plays no role in
the oscillation of magnetization.

(v) A larger $\omega$ would reduce the size of the system. Since the
strength $G$ depends on the spatial wave function and would become
larger if the size is smaller, the effect of interaction would
become stronger if $\omega$ is larger.

(vi) Although the evolution caused by the inclined field does not
depend on $N$ as mentioned, the evolution caused by interaction
does. Since all the solid curves of Fig.~\ref{F1}a appear to be
strictly confined by the dashed curve, how these curves vary with
$N$ depends on how the dashed curve varies with $N$. The latter is
referred to previous literatures\cite{law98}.

(vii) When the quadratic Zeeman term is taken into account, there is
no analytical solution. However, the evolution can be solved
numerically. We found that, when the field is weak (say, $B\leq 0.1\
mG$), the effect of the quadratic term is negligible.

In summary, an analytical approach beyond the mean field theory has
been proposed to describe the spin-evolution of small condensates. A
magnetic field is applied along the $Z^{\prime}$-axis which is
deviated from the $Z$-axis of quantization. Under the
single-mode-approximation, exact time-dependent solution of the
Hamiltonian has been obtained. A formula governing the evolution has
been derived, and related numerical results have been presented. It
was found that a very weak magnetic field can cause a strong, swift,
and nearly periodic oscillation which is highly sensitive to the
magnitude and direction of the magnetic field. The high sensitivity
implies that the phenomenon might be useful for measuring the
direction of a very weak field. When the magnetic field is stronger
and the inclined angle is larger (say, $B$ is in the order of $mG$
and $\theta >\pi /15$), the oscillation cycles will be very dense
and the observation might be misunderstood as random fluctuations.

\begin{acknowledgments}
This work is supported by the NSFC under the grants 10874249 and
from the project of National Basic Research Program of China
(2007CB935500).
\end{acknowledgments}


\begin{thebibliography}{99}

\bibitem{stam98}
 D.M. Stamper-Kurn, M.R. Andrews, A.P. Chikkatur, S. Inouye,
 H.-J. Miesner, J. Stenger, and W. Ketterle,
 Phys. Rev. Lett. \textbf{80}, 2027 (1998).
% Optical Confinement of a Bose-Einstein Condensate

\bibitem{sten98}
 J. Stenger, S. Inouye, D. M. Stamper-Kurn, H. -J. Miesner,
 A. P. Chikkatur, and W. Ketterle,
 Nature (London) \textbf{396}, 345 (1998).
% Spin domains in ground-state Bose-Einstein condensates

\bibitem{ho98}
 T.-L. Ho,
 Phys. Rev. Lett. \textbf{81}, 742 (1998).
% Spinor Bose Condensates in Optical Traps

\bibitem{ohmi98}
 T. Ohmi and K. Machida,
 J. Phys. Soc. Jpn. \textbf{67}, 1822 (1998).
% Bose-Einstein Condensation with Internal Degrees of Freedom in Alkali Atom Gases

\bibitem{goel03}
 A. Gorlitz, T. L. Gustavson, A. E. Leanhardt, R. Low,
 A. P. Chikkatur, S. Gupta, S. Inouye, D. E. Pritchard, and W. Ketterle,
 Phys. Rev. Lett. \textbf{90}, 090401 (2003).
% Sodium Bose-Einstein Condensates in the F=2 State in a Large-Volume Optical Trap

\bibitem{grie05}
 A. Griesmaier, J. Werner, S. Hensler, J. Stuhler, and T. Pfau,
 Phys. Rev. Lett. \textbf{94}, 160401 (2005).
% Bose-Einstein Condensation of Chromium

\bibitem{sore01}
 A. Sorensen, L.-M. Duan, J.I. Cirac, and P. Zoller,
 Nature \textbf{409}, 63 (2001).
% Many-particle entanglement with Bose-Einstein condensates

\bibitem{chang2004}
 M.-S. Chang, C.D. Hamley, M.D. Barrett, J.A. Sauer,
 K.M. Fortier, W.Zhang, L. You, and M.S. Chapman,
 Phys. Rev. Lett. \textbf{92}, 140403 (2004)
% Observation of Spinor Dynamics in Optically Trapped 87Rb Bose-Einstein Condensates

\bibitem{youli2005}
 M.-S. Chang, Q. Qin, W.X. Zhang, L. You, and M.S. Chapman,
 Nature Physics (London) \textbf{1}, 111 (2005).
% Coherent spinor dynamics in a spin-1 Bose condensate

\bibitem{law98}
 C.K. Law, H. Pu, and N.P. Bigelow,
 Phys. Rev. Lett. \textbf{81}, 5257 (1998).
% Quantum Spins Mixing in Spinor Bose-Einstein Condensates

\bibitem{pu99}
 H. Pu, C.K. Law, S. Raghavan, J.H. Eberly, and N.P. Bigelow,
 Rhys. Rev. A. \textbf{60}, 1463 (1999).
% Spin-mixing dynamics of a spinor Bose-Einstein condensate

\bibitem{chan07}
 L. Chang, Q. Zhai, R. Lu, and L. You,
 Phys. Rev. Lett. \textbf{99}, 080402 (2007).
% Number Fluctuation Dynamics of Atomic Spin Mixing inside a Condensate

\bibitem{uchi08}
 S. Uchino, T. Otsuka, and M. Ueda,
 Phys. Rev. A \textbf{78}, 023609 (2008).
% Dynamical symmetry in spinor Bose-Einstein condensates

\bibitem{cui2008}
 X. Cui, Y. Wang, and F. Zhou,
 Phys. Rev. A \textbf{78}, 050701(R) (2008).
% Quantum-fluctuation-driven coherent spin dynamics in small condensates

\bibitem{jing08}
 J. Cheng, H. Jing, and Y. J. Yan,
 Phys. Rev. A \textbf{77}, 061604(R) (2008).
% Spin-mixing dynamics in a spin-1 atomic condensate coupled with a molecular condensate

\bibitem{mark08}
 R. M. Bradley, J. E. Bernard, and L. D. Carr,
 Phys. Rev. A \textbf{77}, 033622 (2008).
% Exact dynamics of multicomponent Bose-Einstein condensates in optical lattices in one, two, and three dimensions

\bibitem{kron08}
 J. Kronj$\ddot{a}$ger, K. Sengstock, K. Bongs,
 New J. Phys. \textbf{10}, 045028 (2008).
% Chaotic dynamics in spinor Bose¨CEinstein condensates

\bibitem{kuwa04}
 T. Kuwamoto, K. Araki, T. Eno, and T. Hirano,
 Phys. Rev. A \textbf{69}, 063604 (2004).
% Magnetic field dependence of the dynamics of 87Rb spin-2 Bose-Einstein condensates

\bibitem{mur06}
 J. Mur-Petit, M. Guilleumas, A. Polls, A. Sanpera, M. Lewenstein,
 K. Bongs and K. Sengstock,
 Phys. Rev. A \textbf{73}, 013629 (2006).
% Dynamics of F=1 87Rb condensates at finite temperatures

\bibitem{kron06}
 J. Kronj$\ddot{a}$ger, C. Becker, P. Navez, K. Bongs, and K. Sengstock,
 Phys. Rev. Lett. \textbf{97}, 110404 (2006).
% Magnetically Tuned Spin Dynamics Resonance

\bibitem{pu00}
 H. Pu, S. Raghavan, and N.P. Bigelow,
 Rhys. Rev. A., \textbf{61}, 023602 (2000).
% Manipulating spinor condensates with magnetic fields: Stochastization, metastability, and dynamical spin localization

\bibitem{yi06}
 S. Yi and H. Pu,
 Phys. Rev. A \textbf{73}, 023602 (2006).
% Magnetization, squeezing, and entanglement in dipolar spin-1 condensates

\bibitem{kemp99}
 E. G. M. van Kempen, S. J. J. M. F. Kokkelmans, D. J. Heinzen,
 and B. J. Verhaar,
 Phys. Rev. Lett. \textbf{88}, 093201 (2002).
% Interisotope Determination of Ultracold Rubidium Interactions from Three High-Precision Experiments

\bibitem{zhan05}
 W. Zhang, D.L. Zhou, M.-S. Chang, M.S. Chapman, and L. You,
 Phys. Rev. A \textbf{72}, 013602 (2005).
% Coherent spin mixing dynamics in a spin-1 atomic condensate

\bibitem{wu96}
 Y. Wu,
 Phys. Rev. A \textbf{54}, 4534 (1996).
% Simple algebraic method to solve a coupled-channel cavity QED model

\bibitem{luo08a}
 M. Luo, C.G. Bao and Z.B. Li,
 J. Phys. B: At. Mol. Opt. Phys. \textbf{41}, 245301 (2008).
% Spin evolution of a mixture of Rb and Na Bose¨CEinstein condensates: an exact approach under the single-mode approximation

\bibitem{bao04}
 C.G. Bao,
 Acta Sci. Nat. Univ. Sunyatseni \textbf{43}, 70 (2004).
% Factional parentage coefficients for spinor Bose-Einstein condensation

\bibitem{bao05}
 C.G. Bao and Z.B. Li,
 Phys. Rev. A \textbf{72}, 043614 (2005).
% First excited band of a spinor Bose-Einstein condensate

\bibitem{luo08b}
 M. Luo, C.G. Bao and Z.B. Li,
 Phys. Rev. A \textbf{77}, 043625 (2008).
% Evolution of the average populations of spin components of spin-1 Bose-Einstein condensates beyond mean-field theory

\end{thebibliography}
\end{document}